\documentclass[10pt]{iopart}

\usepackage{graphicx}
\usepackage{dcolumn}   
\usepackage{bm,amssymb}   
\usepackage{float}
\usepackage{xcolor}
\usepackage{soul}
\usepackage[breaklinks=true,colorlinks=true,linkcolor=blue,urlcolor=blue,citecolor=blue]{hyperref}
\usepackage[normalem]{ulem}

\newcommand{\strikeout}[1]{}

\newcommand{\strikeouts}[1]{}

\begin{document}

\title[Probing non-linear MHD stability of the EDA H-mode in AUG]{Probing non-linear MHD stability of the EDA H-mode in ASDEX Upgrade}

\author{A.~Cathey$^{1}$, M.~Hoelzl$^1$, L.~Gil$^2$, M.G.~Dunne$^1$,  G.F. Harrer$^3$, G.T.A.~Huijsmans$^{4,5}$, J.~Kalis$^1$, K.~Lackner$^1$, S.J.P.~Pamela$^6$, E.~Wolfrum$^1$, S.~G\"unter$^1$, the JOREK team\footnote{see the author list of M. Hoelzl et al 2021 Nucl. Fusion 61 065001}, the ASDEX Upgrade Team\footnote{See the author list of U. Stroth et al 2022 Nucl. Fusion 62 042006}, and the EUROfusion MST1 Team\footnote{see the author list of B. Labit et al. 2019 Nucl. Fusion 59  0860020}} 
\address{$^1$ Max Planck Institute for Plasma Physics, Boltzmannstr.2, 85748 Garching, Germany}
\address{$^2$ Instituto de Plasmas e Fusão Nuclear, Instituto Superior Técnico, Universidade de Lisboa, 1049-001 Lisboa, Portugal}
\address{$^3$ Institute of Applied Physics, TU Wien, 1040 Vienna, Austria}
\address{$^4$ CEA, IRFM, 13108 Saint-Paul-Lez-Durance, France}
\address{$^5$ Eindhoven University of Technology, P.O. Box 513, 5600 MB Eindhoven, The Netherlands}
\address{$^6$ CCFE, Culham Science Centre, Abingdon, Oxon, OX14 3DB, United Kingdom}

\ead{andres.cathey@ipp.mpg.de}

\date{\today}

\begin{abstract}
Regimes of operation in tokamaks that are devoid of large ELMs have to be better understood to extrapolate their applicability to reactor-relevant devices. This paper describes non-linear extended MHD simulations that use an experimental equilibrium from an EDA H-mode in ASDEX Upgrade. Linear ideal MHD analysis indicates that the operational point lies slightly inside of the stable region. The non-linear simulations with the visco-resistive extended MHD code, JOREK, sustain non-axisymmetric perturbations that are linearly most unstable with toroidal mode numbers of ${n=\{6\dots9\}}$, but non-linearly higher and lower $n$ become driven and the low-n become dominant. The poloidal mode velocity during the linear phase is found to correspond to the expected velocity for resistive ballooning modes. The perturbations that exist in the simulations have somewhat smaller poloidal wavenumbers (${k_\theta\sim0.1~\mathrm{to}~0.5~\mathrm{cm^{-1}}}$) than the experimental expectations for the quasi-coherent mode in EDA, and cause non-negligible transport in both the heat and particle channels. In the transition from linear to non-linear phase, the mode frequency chirps down from approximately 35~kHz to 13~kHz, which corresponds approximately to the lower end of frequencies that are typically observed in EDA H-modes in ASDEX Upgrade. 
\end{abstract}

\submitto{\NF}

\maketitle
\ioptwocol

\section{Introduction}

Tokamak operation in the so-called high confinement mode (H-mode) naturally exhibits the onset of repetitive edge localised modes (ELMs), which can expel ${5}$ to ${15~\%}$ of the thermal energy stored in the magnetically confined plasma~\cite{zohm1996edge,leonard2014edge}. For present-day devices, the energy expelled by ELMs is not a cause for concern as it does not exceed the material limits of the plasma facing components. However, for future devices like ITER, which will confine much larger amounts of thermal energy, it is foreseen that large ELMs will not be tolerable~\cite{eich2017elm,Gunn_2017}. Techniques that mitigate or suppress ELMs are actively investigated, but it is presently not possible to determine operational spaces of applicability in future machines. The same problem exists for regimes of operation that are either completely devoid of large ELMs or show only small ELMs that may be tolerable by the plasma facing components. Such no- and small-ELM regimes host some transport mechanism that prevents the edge pressure gradient and current density to grow unconstrained (which constitutes the reason why ELMs become excited) and flushes unwanted impurities out of the confined plasma~\cite{Oyama_2006,Viezzer_2018,Viezzer_2022}.

One specific example of no-ELM regimes is the enhanced D-alpha (EDA) H-mode, which was first observed in the high-field tokamak, Alcator C-mod, with ion-cyclotron heating after a fresh boronisation~\cite{Takase_1997,Greenwald_1999}. The EDA constantly exhibits a quasi-coherent mode (QCM) primarily localised in the low-field side with a radial width of ${\sim2-5~\mathrm{mm}}$ and frequencies of ${50-150~\mathrm{kHz}}$. The QCM which appears to be located in the steep gradient region of the pedestal (at the radial electric field minimum~\cite{Theiler_2017}, but further outside in Ref.~\cite{LaBombard_2014} which probes, and may influence, the plasma). The influence of the QCM is observed in fluctuations of the density, electrostatic potential, and poloidal magnetic field~\cite{Hubbard_2001,Snipes_2001,LaBombard_2014}. As such, the QCM is thought to be an instability with electromagnetic character that constantly regulates the pedestal below the conditions where type-I ELMs are excited by causing continuous particle transport. 

In addition to C-mod, the EDA has also been observed recently in EAST (closely related to a small ELM regime)~\cite{SunPJ_2019}, in ASDEX Upgrade with electron-cyclotron heating (ECRH)~\cite{Gil_2020}, and with a mixture of ECRH and neutral beam injection (NBI), including Argon seeding~\cite{Kallenbach_2021}), and also in DIII-D~\cite{Paz-Soldan_2021}. For the latter two machines,  which have similar sizes and current/magnetic field and are larger than C-mod, the QCM frequency ${f_\mathrm{QCM}=15-40~\mathrm{kHz}}$, and the poloidal wavenumbers range between ${k_\theta=0.5-0.6~\mathrm{cm^{-1}}}$, while in C-mod they range between ${0.8-2.0~\mathrm{cm^{-1}}}$~\cite{Terry_2005}. These differences can be expected due to the different machine sizes; the toroidal mode numbers in question correspond to ${n\approx20}$. In all devices, the QCM moves in the electron diamagnetic direction in the lab frame\footnote{In Alcator C-mod, mode propagation with respect to the plasma frame of reference (i.e., ${v_\mathrm{ExB}}$) has been separately reported in the ion diamagnetic direction for ICRF-heated plasmas as measured with gas puff imaging~\cite{Theiler_2017} and in the electron diamagnetic direction for ohmically-heated plasmas as measured with mirror Langmuir probes~\cite{LaBombard_2014}.} and its frequency starts at larger values (roughly double), chirp down and remain in the aforementioned ranges for several confinement times. Magnetohydrodynamic (MHD) analysis of EDA indicates that the operational points lie slightly inside of the stability boundary for peeling-ballooning modes (the instabilities responsible for ELMs) with medium/high toroidal mode numbers ($n=\{10\dots30\}$) and has been hypothesised that the QCM possibly relates more closely to resistive ballooning modes~\cite{Mossessian_2002,Mazurenko_2002,Myra_2005,Porkolab2006} or to drift ballooning modes~\cite{Hubbard_2001}.

The theoretical framework that relates to the QCM in EDA has not been yet fully explained~\cite{Theiler_2017}. Approaches from modelling include BOUT simulations for Alcator C-mod which connect resistive ballooning modes to the QCM in terms of several distinct features (poloidal wavenumber, frequency, velocity)~\cite{Mazurenko_2002}. BOUT++ has also been applied to study the QCM in C-mod and found good agreement with experiments; such studies suggest that resistive ballooning modes and drift-Alfv\`en waves are the dominant instabilities during the EDA. The gyrokinetic code, GENE, has been used to study an EDA in ASDEX Upgrade, which did not find clear indications of the QCM (possibly due to difficulties encountered in resolving flux surfaces too close to the separatrix)~\cite{Stimmel_2022}. Based on the same
experimental equilibrium on which the GENE simulations had been based, first non-linear extended MHD simulations with the JOREK code~\cite{Hoelzl_2021,huysmans2007mhd} are presented. 

In this paper, non-linear extended MHD simulations with initial conditions chosen from an EDA discharge in ASDEX Upgrade are presented. The description of the experimental discharge and the set-up of the simulations are presented in section~\ref{sec:exp-and-setup}, which includes results from linear ideal MHD calculations for the experimental equilibrium. Thereafter, in section~\ref{sec:lineargrowth} the linearly unstable modes in the
visco-resistive JOREK simulations are described in terms of their structure, growth-rates, mode velocity and direction of propagation and, finally, the resulting non-linear coupling that excites the linearly stable modes. The dominant modes during the non-linear phase (when the non-axisymmetric perturbations interact with the background plasma), the changes caused onto the magnetic fields, and the nature of the fluctuations across the pedestal are detailed in section~\ref{sec:nonlinearphase}.
Finally, section~\ref{sec:discussions-conclusions} summarises and discusses the results presented.

\section{Experimental discharge and simulation set-up}\label{sec:exp-and-setup}

The simulations presented here take their initial conditions from AUG discharge \#36330 in the time range 6.115 to 6.190~s, which is an argon-seeded EDA H-mode first described in Ref.~\cite{Kallenbach_2021}, and that has been simulated with GENE~\cite{Stimmel_2022}. This section presents several details on the experimental discharge as well as newly analysed data from the magnetic pick-up coil signals in section~\ref{ssec:aug36330}, and the JOREK simulation set-up together with relevant details from the model are discussed in section~\ref{ssec:jorek}.

\subsection{AUG \#36330 -- Ar seeded EDA H-mode}\label{ssec:aug36330}

\begin{figure}[!h]
\centering
  \includegraphics[width=0.49\textwidth]{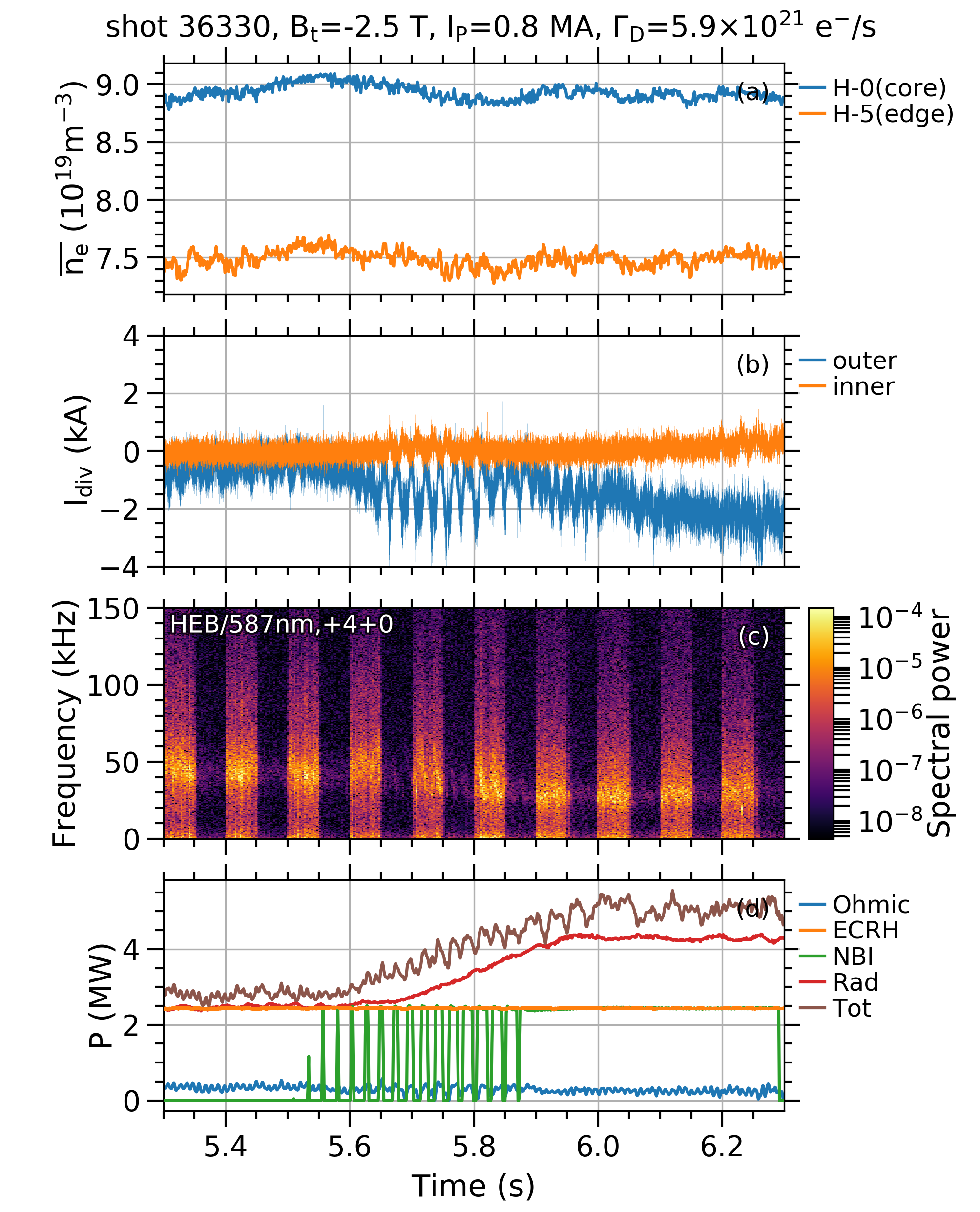}
\caption{Traces of the AUG\#36330 discharge for the time frame 5.3 to 6.3~s. (a) shows the core and edge line-integrated density measurements from interferometry, (b) the inner and outer divertor currents. The frequency spectrogram from a helium beam signal displays a dominant frequency around ${f_\mathrm{QCM}\approx30~\mathrm{kHz}}$ in (c); there are phases without signal because the helium valve is turned on and off periodically in order to subtract the background emission. And the heating power mixture together with the radiated power is shown in (d).}
\label{fig:aug36330-traces}
\end{figure}

The discharge used to set-up the simulations presented in this paper featured a stable EDA H-mode for several seconds. Detailed accounts of this experimental discharge can be found in Refs~\cite{Kallenbach_2021,Stimmel_2022}; here is only a brief summary of the characteristics relevant for the present modelling work. The discharge features a lower single null plasma with elongation ${\kappa=1.6}$ and high triangularity (${\delta_\mathrm{lower}=0.49}$ and ${\delta_\mathrm{upper}= 0.31}$, ${\delta_\mathrm{avg}=0.40}$) operating in favourable ${\nabla B}$ drift direction (where the ion $\bm B \times \nabla B$ drift points towards the active X-point). The toroidal magnetic field is $-2.5~\mathrm{T}$, the plasma current ${0.8~\mathrm{MA}}$, and the edge safety factor ${q_{95}=4.86}$. Figure.~\ref{fig:aug36330-traces} shows the traces of several quantities including the line-averaged density for lines-of-sight that probe the core and the edge (a), the divertor currents (b) where it can be seen that ELMs are not present, a frequency spectrogram from a helium beam signal (c) with a quasi-coherent mode (QCM) with fluctuation frequency of $\sim\{20\dots40\}~\mathrm{kHz}$, and the heating mix (d). The latter is comprised of ECRH and NBI with ${2.5~\mathrm{MW}}$ each from 5.9 to 6.3~s. The JOREK simulations are initialised from a time point within this phase. 

The absence of ELMs observed in fig.~\ref{fig:aug36330-traces}(b) is likely attributed to the electromagnetic QCM with ${f_\mathrm{QCM}\approx30~\mathrm{kHz}}$ (ranges between ${20-40~\mathrm{kHz}}$), which can be observed in fig.~\ref{fig:aug36330-traces}(c). The transport caused by the cross-field transport, including the QCM, causes a peak heat-flux onto the divertor targets of ${5-10~\mathrm{MW/m^2}}$ (derived from Langmuir probes)~\cite{Kallenbach_2021}. 

\begin{figure}[!ht]
\centering
  \includegraphics[width=0.45\textwidth]{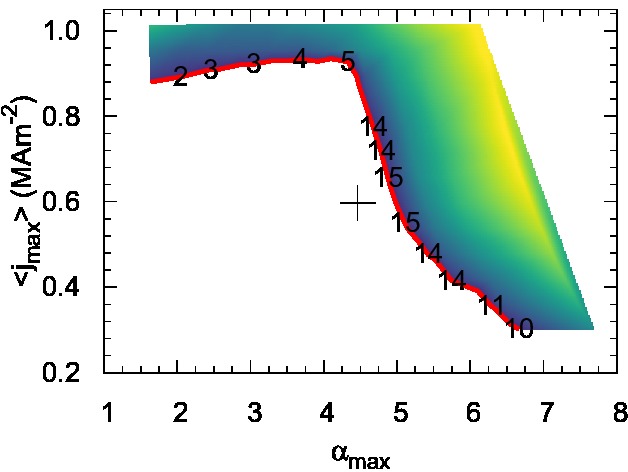}
\caption{MISHKA ideal linear MHD analysis of the equilibrium reconstruction at ${6.165~\mathrm{s}}$. The operational point at ${6.165~\mathrm{s}}$ is found to be just inside of the ideal ballooning (${n\approx15}$) boundary.}
\label{fig:mishka}
\end{figure}

The high-resolution equilibrium reconstruction with bootstrap constraint is generated with CLISTE~\cite{mc1999analytical} from data in the time range 6.115 to 6.190~s, and it is used to initialise the JOREK simulations that will be discussed in depth in the following sections. Linear ideal MHD stability analysis with MISHKA~\cite{Mikhailovskii_1997} is performed for this equilibrium, and shown in fig.~\ref{fig:mishka}. The linear stability analysis finds the operational point to be slightly inside of the (${n\approx15}$) ballooning boundary. However, the MISHKA simulations neglect all non-ideal effects, notably the destabilizing resistivity and the stabilizing ExB and diamagnetic flows. Non-linear~\cite{Cathey_2022,Kleiner_2021} and linear~\cite{Nystrom_2022} resistive MHD simulations have shown to move the ballooning boundary to lower pressure gradients, such that the operational point could easily be unstable to resistive (peeling-)ballooning modes. Indeed, the JOREK simulations that will be presented in the following sections find that the equilibrium is in fact unstable to resistive peeling-ballooning modes at realistic plasma resistivity. 

\begin{figure}[!h]
\centering
  \includegraphics[width=0.45\textwidth]{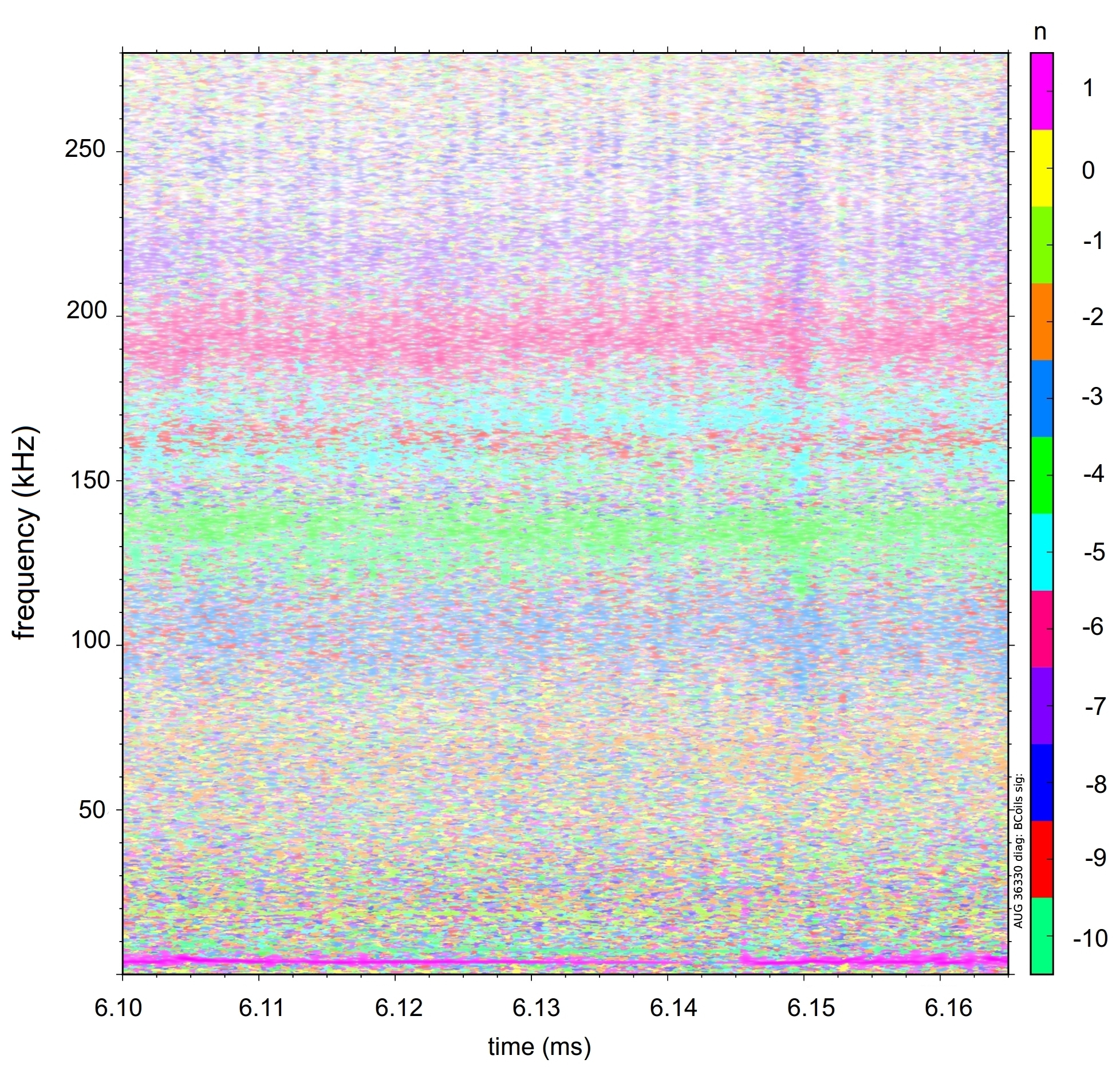}
\caption{Cross phaseogram of several signals from magnetic pick-up coils in the equatorial midplane at different toroidal angles. Different colours indicate distinct toroidal mode numbers. In particular, $n=3\dots7$ are observed to be the relevant toroidal harmonics.}
\label{fig:aug36330-coils}
\end{figure}

In addition to the QCM fluctuations that are observed in the helium beam diagnostic (which can also be observed in spectrograms of reflectometry, interferometry, electron cyclotron emission, magnetic pick-up coils closest to the plasma~\cite{Gil_2020}), magnetic fluctuations are measured during the EDA H-mode phase. By considering several magnetic pick-up coils in the equatorial midplane, but at different toroidal angles, it is possible to identify the relevant toroidal mode numbers of ${n=\{3\dots7\}}$. The cross-phaseogram that allows such analysis is shown in fig.~\ref{fig:aug36330-coils}, but the precise nature of these modes, and their relation (if any) to the QCM is unclear as of yet. For this particular case, the frequencies of the bands with distinct toroidal mode numbers can be written as ${f_{n} = n \times 32.9~\mathrm{kHz}}$. However, it must be noted that in several EDA discharges which display visible bands in the magnetic pick-up coils cross-phaseogram the difference between the base frequency with the QCM frequency can be ${\lesssim 15~\mathrm{kHz}}$.

\subsection{JOREK simulation set-up}\label{ssec:jorek}

As mentioned before, the JOREK simulations are initialised from an experimental reconstruction corresponding to ${6.165~\mathrm{s}}$ in AUG discharge~\#36330. The corresponding simulations are carried out with the single-temperature reduced MHD model within JOREK and considers a perfectly conducting wall as boundary condition. Note that reduced MHD can be applied here allowing to reduce computational costs, since excellent agreement of reduced and full MHD models for pedestal applications has been shown~\cite{Pamela_2020}. For all simulations, the diamagnetic drift and bootstrap current source extensions are considered. The latter means that when profiles change non-linearly, the bootstrap current fraction changes according to the Sauter analytical model~\cite{sauter1999neoclassical,sauter2002erratum}; if profiles remain stationary, the bootstrap current fraction remains stationary as well. Using the former means that the radial electric field well, which is characteristic of experimental $E_r$ profiles in H-mode and proportional to ${\nabla p/n}$, is present in the simulations as has been shown in previous work~\cite{Cathey_2021}. A detailed account of the JOREK code, and of the single temperature reduced MHD model can be found in Ref.~\cite{Hoelzl_2021}.

\begin{figure}[!ht]
\centering
  \includegraphics[width=0.45\textwidth]{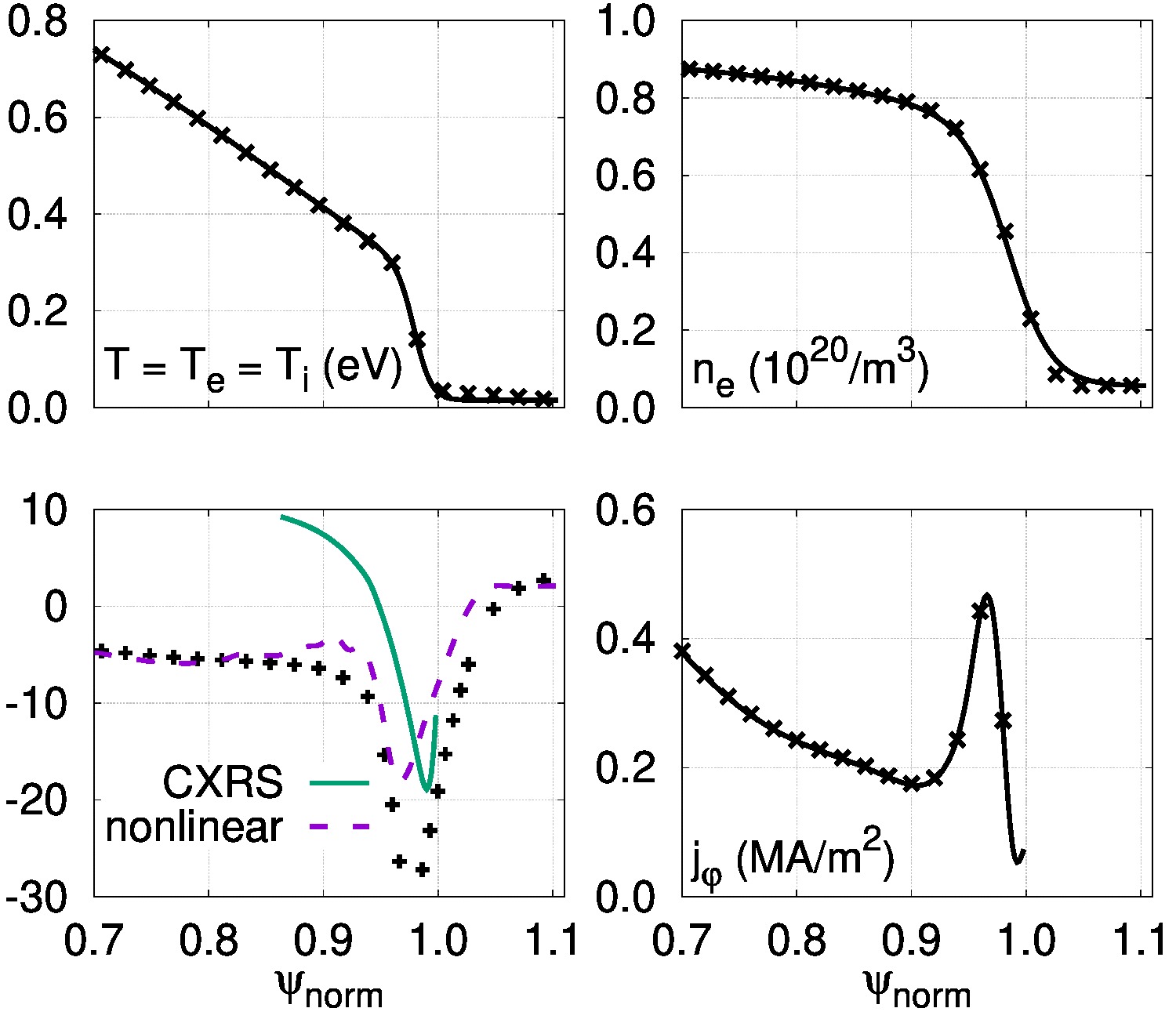}
\caption{The input profiles of density, temperature, radial electric field, and toroidal current density. The latter is comprised of ohmic and bootstrap contributions and it is a flux-surface averaged profile, while the rest are profiles at the outboard midplane. Full black lines are the initial profiles and black crosses represent profiles after $10~\mathrm{ms}$ of axisymmetric simulation. A profile fitted from CXRS measurements for the $E_r$ is shown in full green line from ${\psi_\mathrm{norm}=0.85}$ to 1.00, and a toroidally-averaged midplane profile from a non-linear simulation (${n=0\dots13}$) in a dashed purple line.}
\label{fig:aug36330-profiles}
\end{figure}

The input profiles of the density, temperature, radial electric field, and toroidal current density at the outboard midplane (flux-surface average for ${j_\varphi}$) are shown in fig.~\ref{fig:aug36330-profiles} in full black lines. Black crosses represent profiles after ${10~\mathrm{ms}}$ of axisymmetric evolution (i.e., in the absence of perturbations) are shown. The radial electric field profile in full green line (which only spans until 1.00) comes from charge exchange recombination spectroscopy measurements for the impurity toroidal and poloidal velocity of and the force balance equation ${E_r=\nabla p_\alpha/(e Z_\alpha n_\alpha) - \bm v_\alpha \times \bm B}$, where $\alpha$ is the ${\mathrm{N}^{7+}}$ impurity species. The uncertainty of the $E_r$ minimum is roughly ${\sim\pm4~\mathrm{kV/m}}$~\cite{Stimmel_2022}. It is evident that the axisymmetric profile does not match closely to the measured profile, which is partly because the present simulations do not include a source for the toroidal rotation caused by the NBI torque and partly because the single temperature treatment cannot account for lower ion pressure gradients than electron and drives $E_r$ to be more negative in the pedestal middle. However, when the non-axisymmetric simulation (${n=0\dots13}$) develops (dashed purple line) the minimum value of $E_r$ matches with the experimental measurements due to modifications to the profiles caused by the perturbations which will be described in detail in section~\ref{sec:nonlinearphase}. 

The axisymmetric evolution of the profiles is determined by diffusion coefficients and sources of particles and heat, which were designed to maintain the initial temperature, density, and current density profiles constant, as depicted by the fact that the full black lines and the black crosses in fig.~\ref{fig:aug36330-profiles} overlap. The physical meaning of the diffusion coefficients corresponds to neoclassical and turbulent transport that cannot be simulated with JOREK, but is included with these \textit{ad-hoc} profiles.

The grid resolution required to resolve the simulations comprises 214 in the radial direction (i.e., from the axis to the perfectly conducting wall) which are accumulated primarily in the pedestal region such that the radial extent of one grid element is ${1.28~\mathrm{mm}}$, 354 in the poloidal direction, and 64 toroidal planes, which are needed to include $14$ toroidal Fourier harmonics in the simulation: ${n=0\dots13}$. These parameters make it possible to include a realistic parallel-to-perpendicular heat diffusion anisotropy of the order ${10^{9}}$. The temporal resolution in JOREK is not constrained by the CFL criterion since an implicit time stepping scheme is used; the time step used for the simulations corresponds to ${\sqrt{\mu_0 \rho_0}=0.6687~\mathrm{\mu s}}$ (except in the non-linear phase when it is sometimes necessary to decrease it by half to improve convergence of the iterative solver). Since the MHD model is valid for perturbations of low frequency (relative to the ion cyclotron frequency, ${\Omega_i=e B /m \sim 10^{8}~\mathrm{1/s}}$), perturbations that are faster cannot be adequately resolved with JOREK even if the time step were set to ${\delta t\lesssim1/\Omega_i\sim0.01\sqrt{\mu_0 \rho_0}}$.

Having described the experimental discharge, the simulation set-up and initial conditions, the focus is turned to describing the results during the linear phase. 

\section{Linear growth phase and dependencies}\label{sec:lineargrowth}

The present section describes the simulation results during the linear growth phase including linear growth rates, and the dynamics of the non-axisymmetric perturbations during the early non-linear phase, where three-wave interactions allow linearly stable modes to become non-linearly excited. The location and velocity of the modes during the linear phase is further discussed and their relation to resistive ballooning modes is highlighted.

\subsection{Growth rates and mode localisation}

\begin{figure}[!ht]
\centering
  \includegraphics[width=0.45\textwidth]{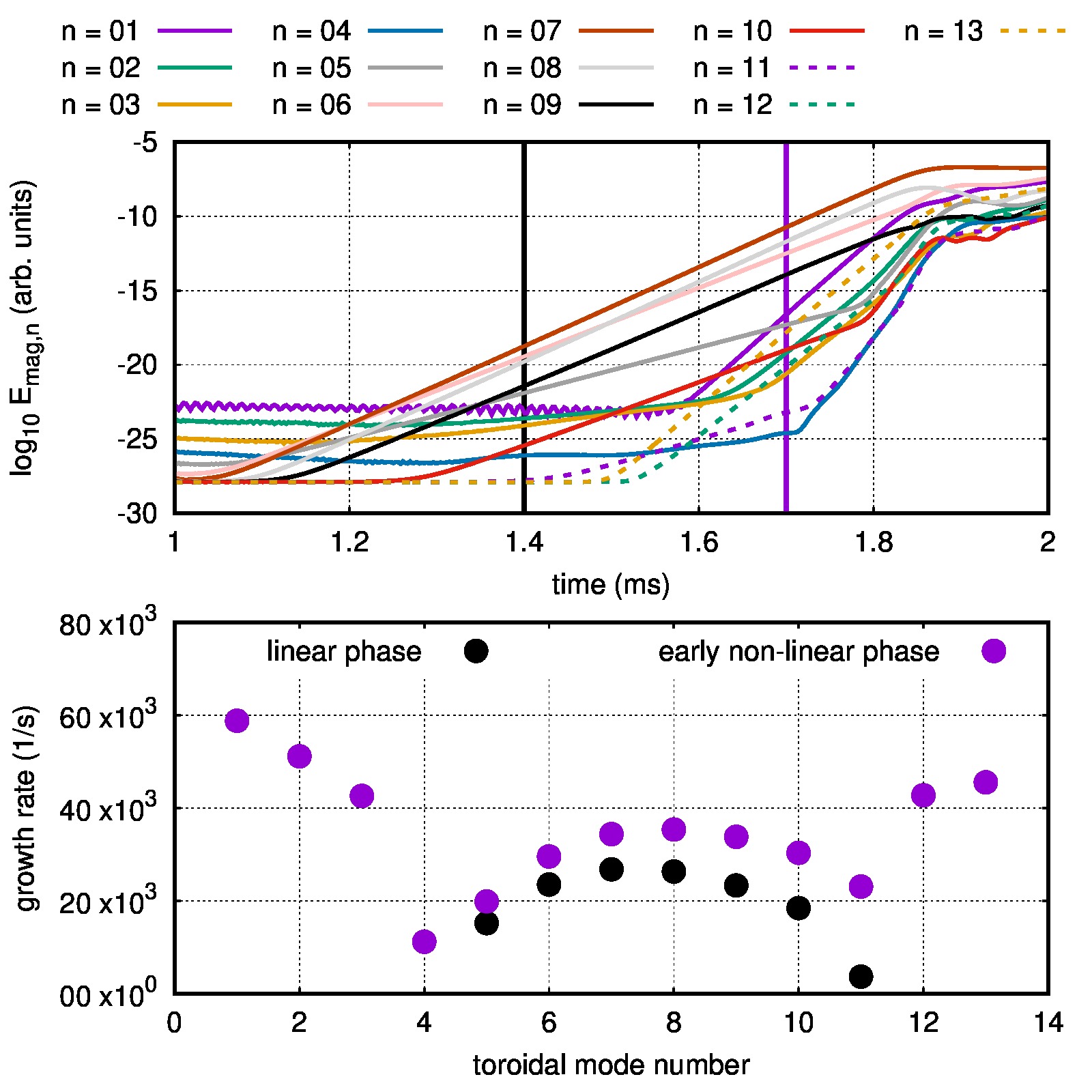}
\caption{Top: magnetic energies of the non-axisymmetric perturbations during the linear phase (${\lesssim1.5~\mathrm{ms}}$) and early non-linear phase (${\gtrsim1.5~\mathrm{ms}}$). Bottom: growth-rates for the different toroidal mode numbers during the linear phase (black squares) and during the early non-linear phase (purple circles). Linearly stable modes become non-linearly destabilised due to three-wave interactions.}
\label{fig:base-Emag-Gmag}
\end{figure}

After ${1~\mathrm{ms}}$ of axisymmetric simulation (during this time the parallel and poloidal flows establish and saturate), non-axisymmetric perturbations with toroidal Fourier harmonics ${n=1\dots13}$ are initialised at noise-amplitude. Figure ~\ref{fig:base-Emag-Gmag} shows the non-axisymmetric perturbation magnetic energies (top) and their growth rates during the linear and early non-linear phases (bottom). It can be distinguished that there are linearly unstable modes (${n=\{5\dots11\}}$) and linearly stable modes (${n=\{1\dots4\},~12}$, and 13) that become non-linearly excited during the early non-linear phase through three-wave interactions. Namely, a mode $n_3$ is driven by $n_1$ and $n_2$ provided that ${n_3=n_1\pm n_2}$ and the resulting growth-rate corresponds to ${\gamma_{n_3} = \gamma_{n_1}+\gamma_{n_2}}$~\cite{krebs2013nonlinear}. In this case several pairs of modes can contribute to driving a linearly stable mode. For example, ${n_1=6}$ and ${n_2=6}$ drive ${n_3=12}$, which is also driven by ${n_1=5}$ and ${n_2=7}$ and any other combination that satisfy ${n_3=n_1\pm n_2}$. The resulting growth-rate of $n=12$ is given by contributions from all pairs of modes that satisfy ${n_1 \pm n_2 = 12}$, but it is mostly determined by the dominant mode pair that drive it.

\subsection{Mode location and velocity}

\begin{figure}[!ht]
\centering
  \includegraphics[width=0.45\textwidth]{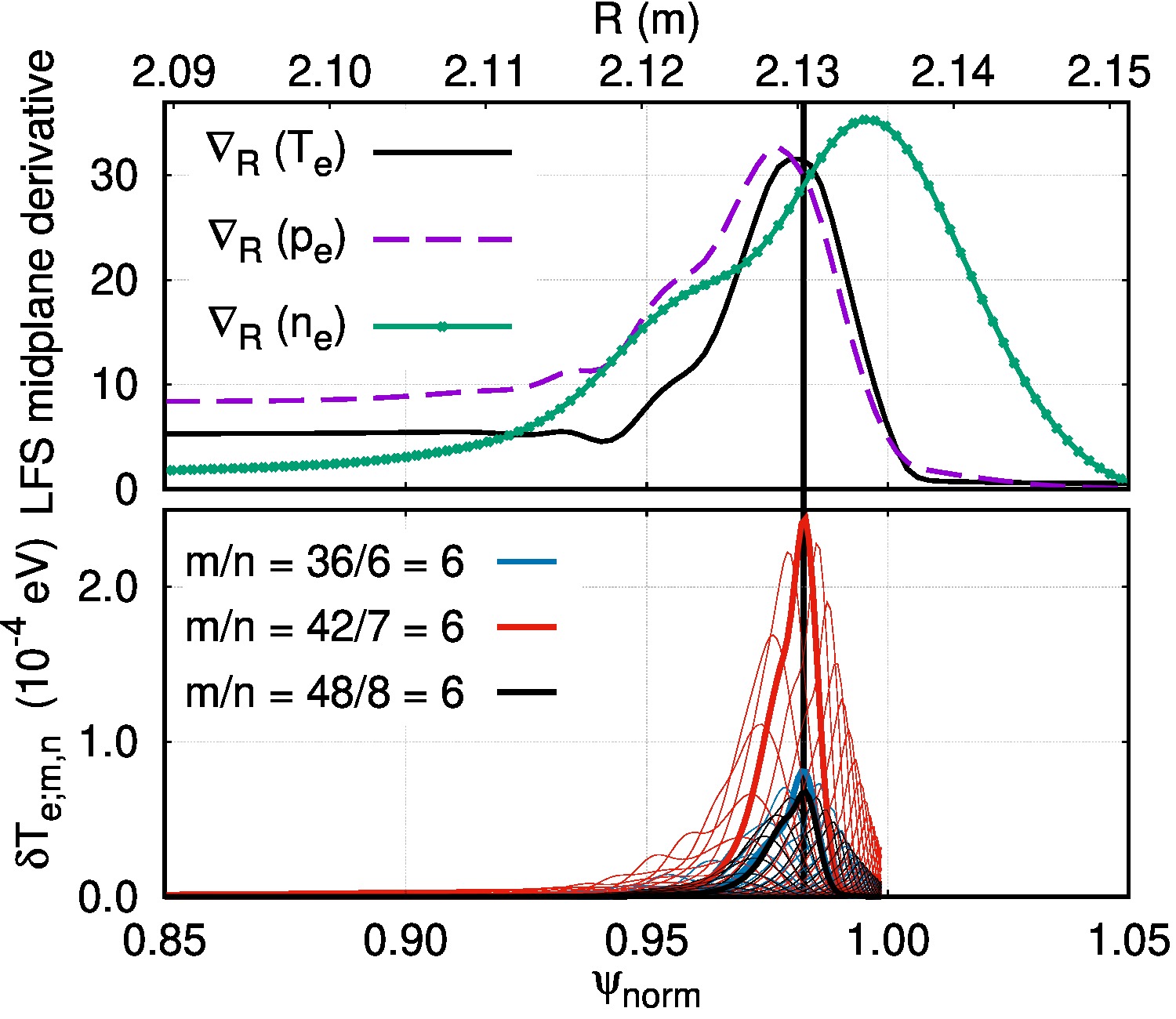}
\caption{Top: derivatives (with respect to the major radius) at the outer midplane for $T_e$, $p_e$, and $n_e$ profiles; the maximum gradient is located at ${\psi_\mathrm{norm}=0.980}$, 0.977, and 0.995, respectively. Bottom: absolute value of various $m/n$ Fourier coefficients for ${n=6}$ in blue, ${7}$ in red, and $8$ in black. The respective thick lines show the ${m/n=36/6}$, $42/6$, and $48/6$; namely at the $q=6$ rational surface.}
\label{fig:base-four2d}
\end{figure}

The location of the linearly unstable modes is around ${\psi_\mathrm{norm}\approx0.98}$, which corresponds to the maximum temperature gradient and is consistent with experimental measurements from Alcator C-mod~\cite{Theiler_2017}, and with recent observations from the EDA in ASDEX Upgrade with Helium beam~\cite{Kalis_private}. To clearly illustrate this, fig,~\ref{fig:base-four2d} (top) shows the derivatives of the electron temperature, pressure, and density with full black line, dashed purple line, and green line with marks, respectively. And fig.~\ref{fig:base-four2d} (bottom) shows the absolute value of the dominant complex ${m/n}$ Fourier coefficients in blue for ${n=6}$, red for $7$, and black for $8$. With thicker lines, the resonant surface of ${q=6}$ is shown for each harmonic with the respective colours. Unsurprisingly, the dominant poloidal mode number during the linear phase is ${m=\{35\dots56\}}$. The location of the mode maximum is shown with a vertical black line that spans both plots and corresponds to ${\psi_\mathrm{norm}=0.983}$, which sits almost exactly at the ${\nabla_R (T_e)}$ maximum.

\begin{figure}[!ht]
\centering
  \includegraphics[width=0.45\textwidth]{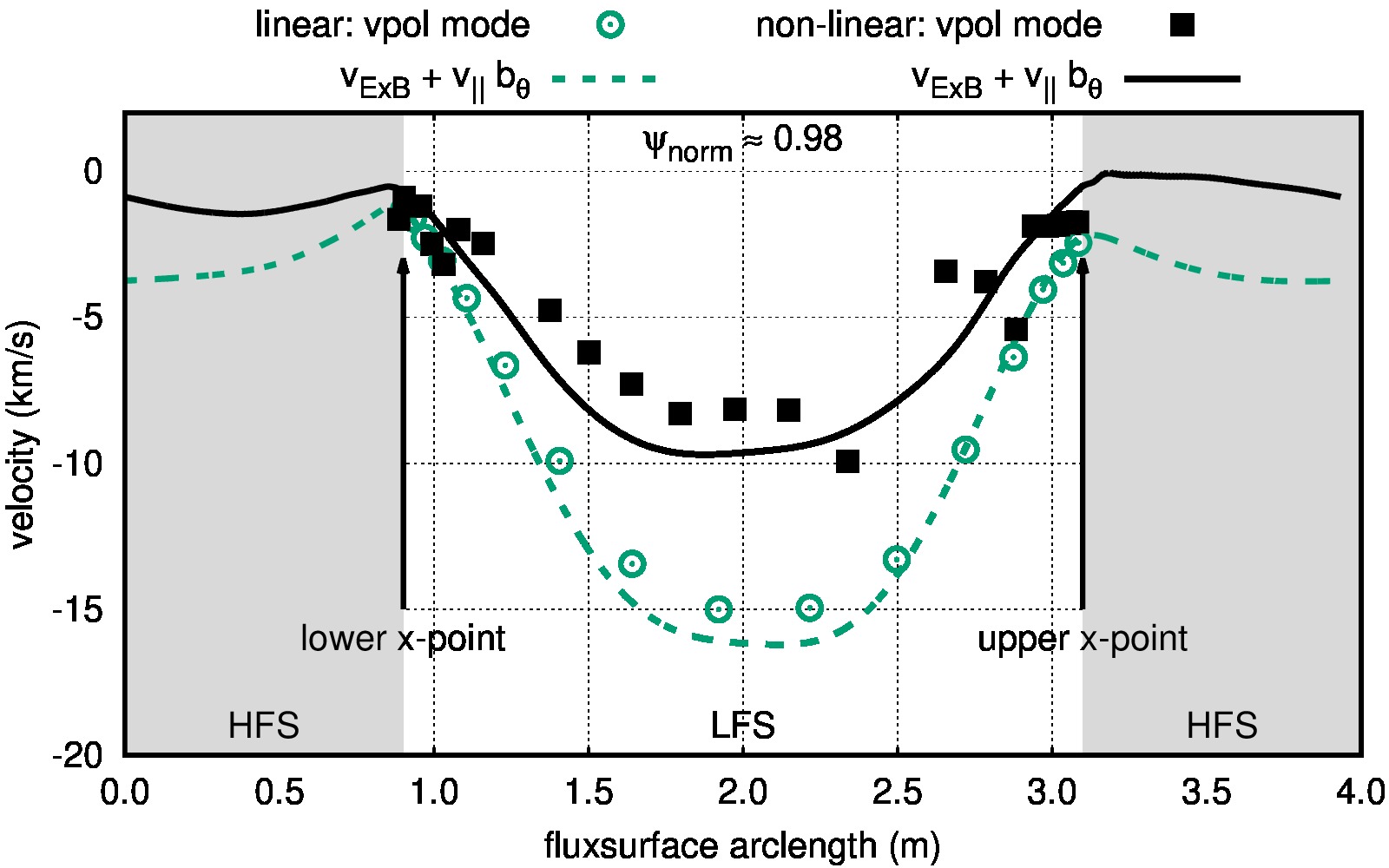}
\caption{Poloidal velocity of the non-axisymmetric perturbations during the linear phase (at ${t=1.60~\mathrm{ms}}$) and non-linear phase (${t=2.70~\mathrm{ms}}$) together with the poloidal velocity of the plasma along the ${\psi_\mathrm{norm}\approx0.98}$ flux-surface.}
\label{fig:base-vmode}
\end{figure}

Figure~\ref{fig:base-vmode} shows the poloidal velocity of the perturbations along the flux-surface where the modes are located (${\psi_\mathrm{norm}\approx0.980}$) during the linear phase (${t=1.60~\mathrm{ms}}$) with green circles and during the non-linear phase (${t=2.70~\mathrm{ms}}$) with black squares. Additionally, the plasma rotation velocity (${v_\mathrm{ExB} + \bm v_{\parallel}\cdot \bm b_\theta}$) is also shown for either phase with a green dashed line and a black full line, respectively. The reduction in mode velocity between linear and non-linear phases will be discussed further in section~\ref{sec:nonlinearphase}. The movement of the non-axisymmetric perturbation is tracked along individual flux-surfaces and the distance travelled is used to obtain the poloidal mode velocity. From ${\psi_\mathrm{norm}=0.929}$ until ${0.987}$ (which is the outermost flux-surface with a reliable measurement of the velocity), the mode velocity at the outboard midplane is roughly ${-15~\mathrm{km/s}}$, where the negative sign indicates movement in the electron diamagnetic drift direction. In the laboratory frame, the modes travel in the electron diamagnetic direction (negative velocities in fig.~\ref{fig:base-vmode}) at all radial locations. In the plasma frame (i.e., the difference between the green circles and dashed line) the modes move in the ion diamagnetic direction at the outboard midplane with ${v_{\mathrm{mode,pl.}}\approx1~\mathrm{km/s}}$ only close to the location where the mode amplitude is maximised (from ${\psi_\mathrm{norm}\approx0.970}$ until $0.982$). Inward and outward of these locations, the modes move in the electron diamagnetic direction in the plasma frame. The mode velocity in the plasma frame is faster in radial locations further away from ${\psi_\mathrm{norm}\approx0.980}$ since the modes experience rigid body rotation. For instance, ${v_\mathrm{mode,pl.}(\psi_\mathrm{norm}=0.934) = -9~\mathrm{km/s}}$ and ${-4.5~\mathrm{km/s}}$ at ${0.997}$, but where the mode amplitude is maximised at ${0.980}$, as mentioned before, it is approximately ${+1~\mathrm{km/s}}$. 

Experimental measurements from Alcator C-mod have measured the mode velocity in the laboratory frame to be in the electron diamagnetic direction always. In ohmic heated-only EDA H-modes, it has been reported that the QCM moves in the electron diamagnetic direction in the plasma frame (with measurements with mirror Langmuir probes)~\cite{LaBombard_2014} and in ICRF-heated EDA in the ion diamagnetic direction in the plasma frame (with measurements from gas puff imaging)~\cite{Theiler_2017}. The latter showed that the mode amplitude is maximised in the maximum electron temperature gradient and in that location ${v_\mathrm{mode,pl.}}$ is measured to be in the ion diamagnetic direction, but at larger velocities than observed in the simulations in the range ${\psi_\mathrm{norm}=\{0.95\dots0.99\})}$, the modes move with ${\{+15\dots+7\}~\mathrm{km/s}}$ (in these experiments, the mode velocity is compared with the ExB velocity, but the poloidal projection of the parallel velocity is neglected). At this stage it is worth mentioning that the simplified scrape-off layer model used in the single-temperature JOREK simulation results in a radial electric field that is unrealistic in the SOL. Future work may then involve a more advance SOL model~\cite{Korving_2023} and a separation between ion and electron temperatures to understand the influence onto the mode dynamics, but this goes beyond the scope of the present work.

From a ballooning mode dispersion relation, the velocity of ideal ballooning modes has been reported in Ref.~\cite{morales2016edge} to be ${v_{\mathrm{ExB}} + \bm v_{\parallel}\cdot \bm b_\theta + v^*_{i,\theta}/2}$, where ${v_\mathrm{ExB}}$ lies solely in the poloidal direction (this is one of the assumptions for the reduced MHD model in JOREK~\cite{Cathey_2021}), and for resistive ballooning modes to be ${v_{\mathrm{ExB}} + \bm v_{\parallel}\cdot \bm b_\theta}$. The mode velocity extracted from the simulations is then compared to the local ${v_\theta = \left(\bm v_{\mathrm{ExB}} + \bm v_{\parallel}\right) \cdot \bm b_\theta}$, which varies across and along flux-surfaces. The mode velocity is found to match best with the local ${v_\theta}$ where the mode amplitude is largest (i.e., ${\psi_\mathrm{norm}\approx0.980}$), as mentioned before. This is an indication that the relevant modes are resistive peeling-ballooning modes. 

\begin{figure}[!ht]
\centering
  \includegraphics[width=0.45\textwidth]{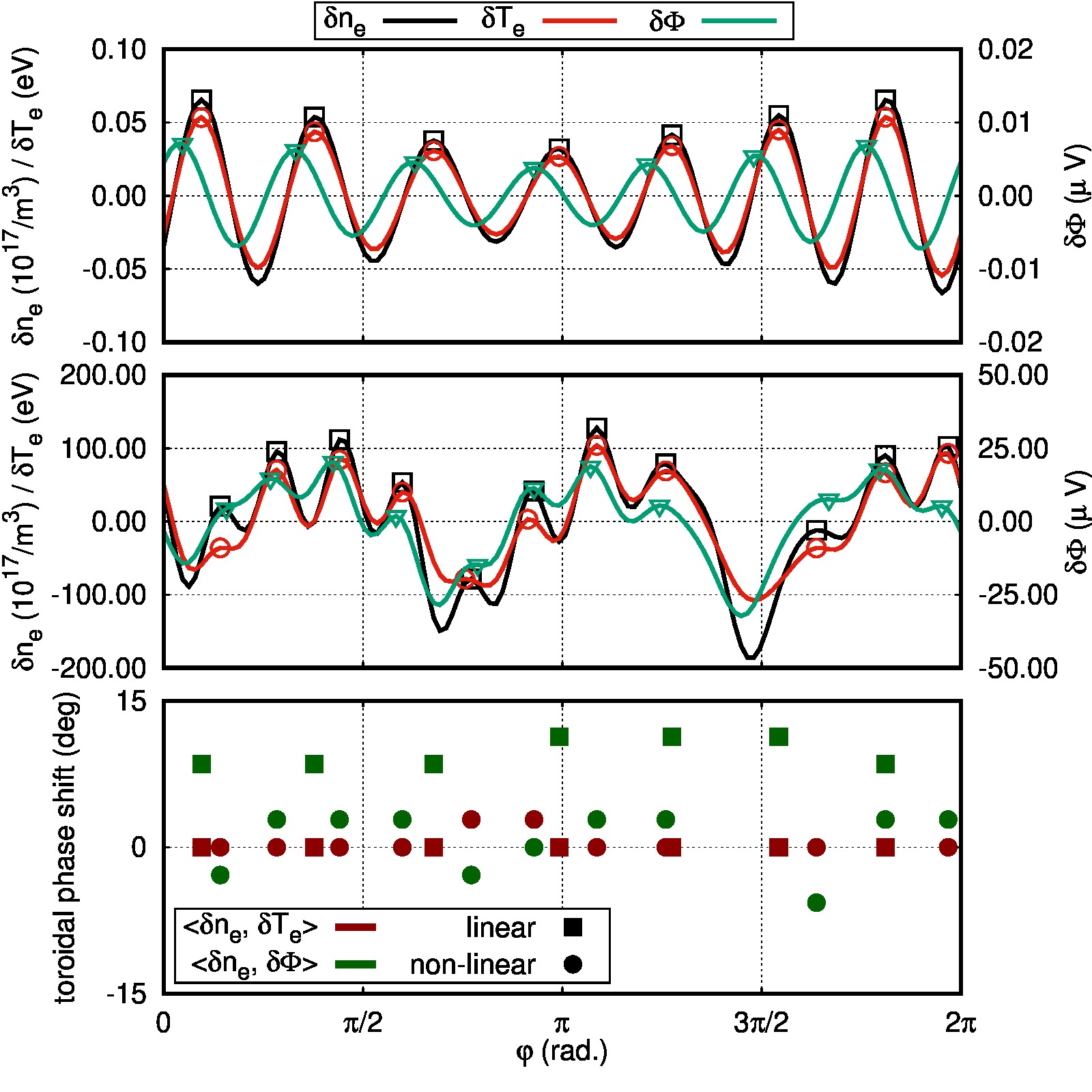}
\caption{Perturbations of density, temperature, and electrostatic potential during the linear phase (top), during the non-linear phase (middle) along a toroidal line with constant $(R,Z_\mathrm{axis})=(2.130,0.042)~\mathrm{m}$. This location is at the outer midplane roughly where the temperature gradient is maximised. The cross-phases ${\langle n_e, T_e \rangle}$ in dark-red and ${\langle n_e, \Phi \rangle}$ in dark-green (bottom) are shown with squares for the linear phase and circles for the non-linear phase.}
\label{fig:base-linear-toroidal-pert}
\end{figure}

The nature of the $n=7$-dominated perturbation along a toroidal line at ${(R,Z)=(2.130,0.042)~\mathrm{m}}$, which is roughly at the maximum temperature gradient location (${\psi_\mathrm{norm}=0.98}$), is shown in fig.~\ref{fig:base-linear-toroidal-pert}. It shows variations of the density, temperature, and electrostatic potential during the linear phase (top) and during the non-linear phase (middle, and further explained in section~\ref{sec:nonlinearphase}). The cross-phase in the toroidal angle, $\varphi$, for ${\langle n_e, T_e \rangle}$ in dark-red and for ${\langle n_e, \Phi \rangle}$ in dark-green are shown for the linear phase with squares and for the non-linear phase with circles (bottom). During the linear phase, the perturbations of density and temperature are `perfectly' in-phase with each-other, but have a cross-phase of ${\sim10^\circ}$ with the electrostatic potential. 

During the linear and early non-linear phase, where the non-axisymmetric perturbations do not cause changes to the axisymmetric background, the perturbations are well described with a fluctuation frequency of ${\sim35~\mathrm{kHz}}$. The frequency of the quasi-coherent mode in the experimental discharge can be observed in fig.~\ref{fig:aug36330-traces} (d), and it is clear that the fluctuation present in the simulation during the linear phase corresponds well with the experimental $f_\mathrm{QCM}$. It is worth reiterating at this stage that during the linear phase the $E_r$ minimum in the simulations reaches ${-29~\mathrm{kV/m}}$, while in the experiments it is calculated to be ${-20\pm4~\mathrm{kv/m}}$. 

\section{Non-linear simulation results}\label{sec:nonlinearphase}

The previous section described the linear phase in terms of growth-rates, mode structure (dominant toroidal and poloidal mode numbers), location, and velocity. The present section details the temporal dynamics at play when the non-axisymmetric perturbations become large enough to interact with the axisymmetric background plasma. The non-linear phase observes a shift of dominant mode numbers from higher (${n=\{6\dots9\}}$) to lower (${n=\{2\dots5\}}$). The mode activity causes perturbations to the background magnetic field strong enough to create an ergodised layer in the pedestal (outward of ${\psi_\mathrm{norm}\approx0.965}$) and enhance parallel heat transport, which results in a depletion of the temperature pedestal. In addition, convective cells are formed and particle transport is generated which also causes a reduction of the pedestal top density. 

\subsection{Shifting dominant toroidal mode numbers}

\begin{figure}[!ht]
\centering
  \includegraphics[width=0.45\textwidth]{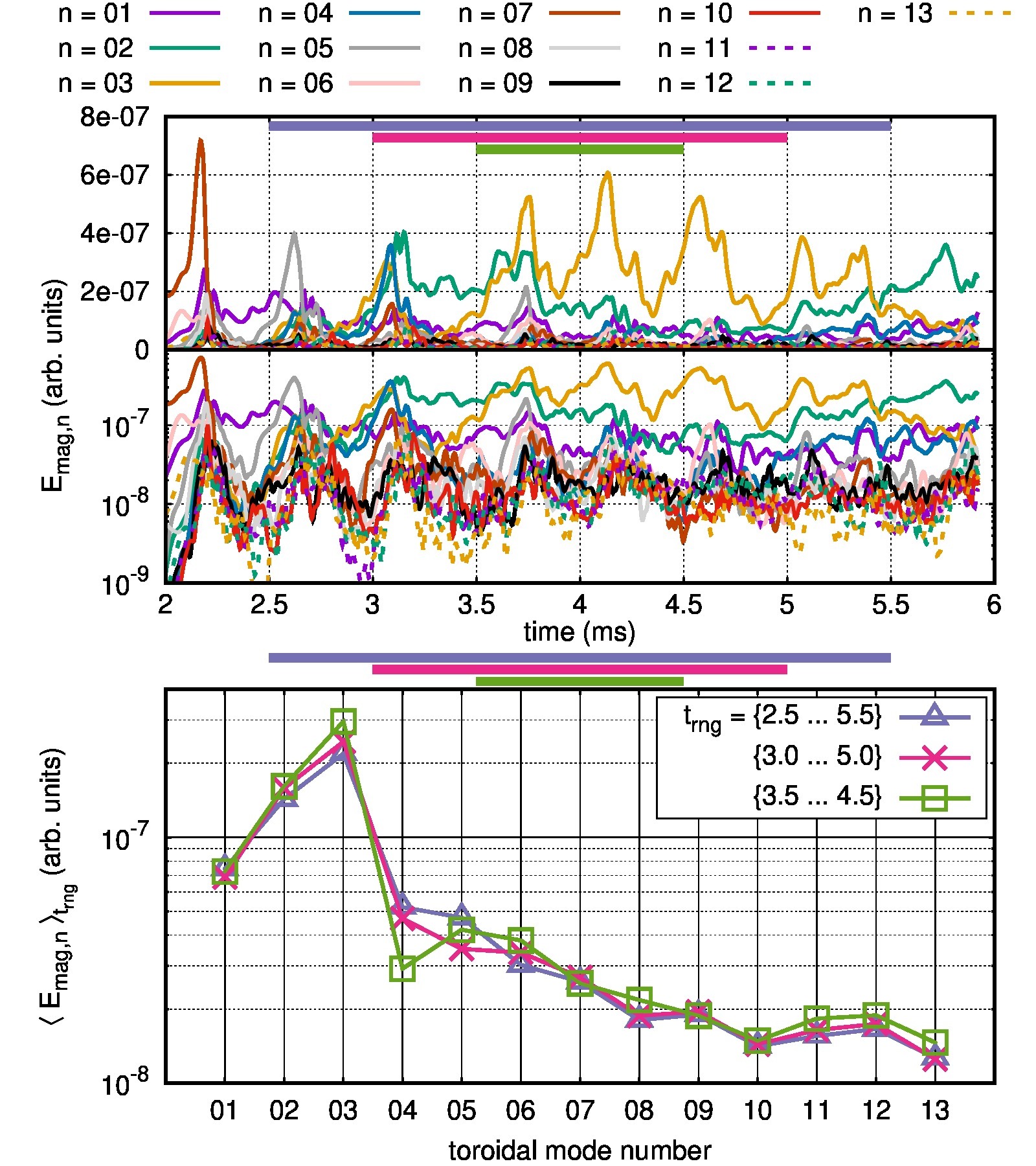}
\caption{Linear (top) and logarithmic (middle) magnetic energies of the non-axisymmetric perturbations during the non-linear phase (from 2 to ${6~\mathrm{ms}}$). Bottom: three different temporally-averaged spectra centered around ${4~\mathrm{ms}}$ with sample sizes of 1, 2, and ${3~\mathrm{ms}}$.}
\label{fig:base-nonlinear-energies}
\end{figure}

The non-linear evolution of the non-axisymmetric perturbations (their magnetic energies) is shown in fig.~\ref{fig:base-nonlinear-energies} from 2 to ${6~\mathrm{ms}}$ in linear (top) and logarithmic (bottom) scales, and three different temporally-averaged spectra (considering sample sizes of 1, 2, and ${3~\mathrm{ms}}$) centered around ${4~\mathrm{ms}}$ (bottom). From the first two figures it is clear that the dominant mode numbers shift from higher-to-lower in time. There are two reasons for this effect 1) non-linear mode coupling gives energy from the higher-n perturbations that are linearly unstable (${n=\{5\dots11\}}$, but in particular ${n=6,}$ 7, and 8) to the lower-n that are linearly stable (${n=\{1\dots4\}}$) and 2) once the linearly most-unstable modes start to have an impact on the axisymmetric background they cause the pedestal density and temperature to become depleted, which in itself always has an effect on the linear spectrum. 

Figure~\ref{fig:base-nonlinear-energies} (bottom) clearly shows that once the non-linear phase is underway the dominant modes become ${n=2}$ and 3. The ${n=3}$-dominated structure can be distinguished in fig.~\ref{fig:base-linear-toroidal-pert} (middle), where also several (11) local maxima can be observed. The presence of this $n=11$ structure is a clear indication that high toroidal mode numbers are involved in the non-linear dynamics as well. During the non-linear phase, density, temperature, and electrostatic potential are roughly in-phase (circles in the bottom part of fig.~\ref{fig:base-linear-toroidal-pert}). The effect of the non-axisymmetric perturbations onto the background plasma is detailed in the following. 

\subsection{Interaction with the background plasma}

\begin{figure}[!ht]
\centering
  \includegraphics[width=0.45\textwidth]{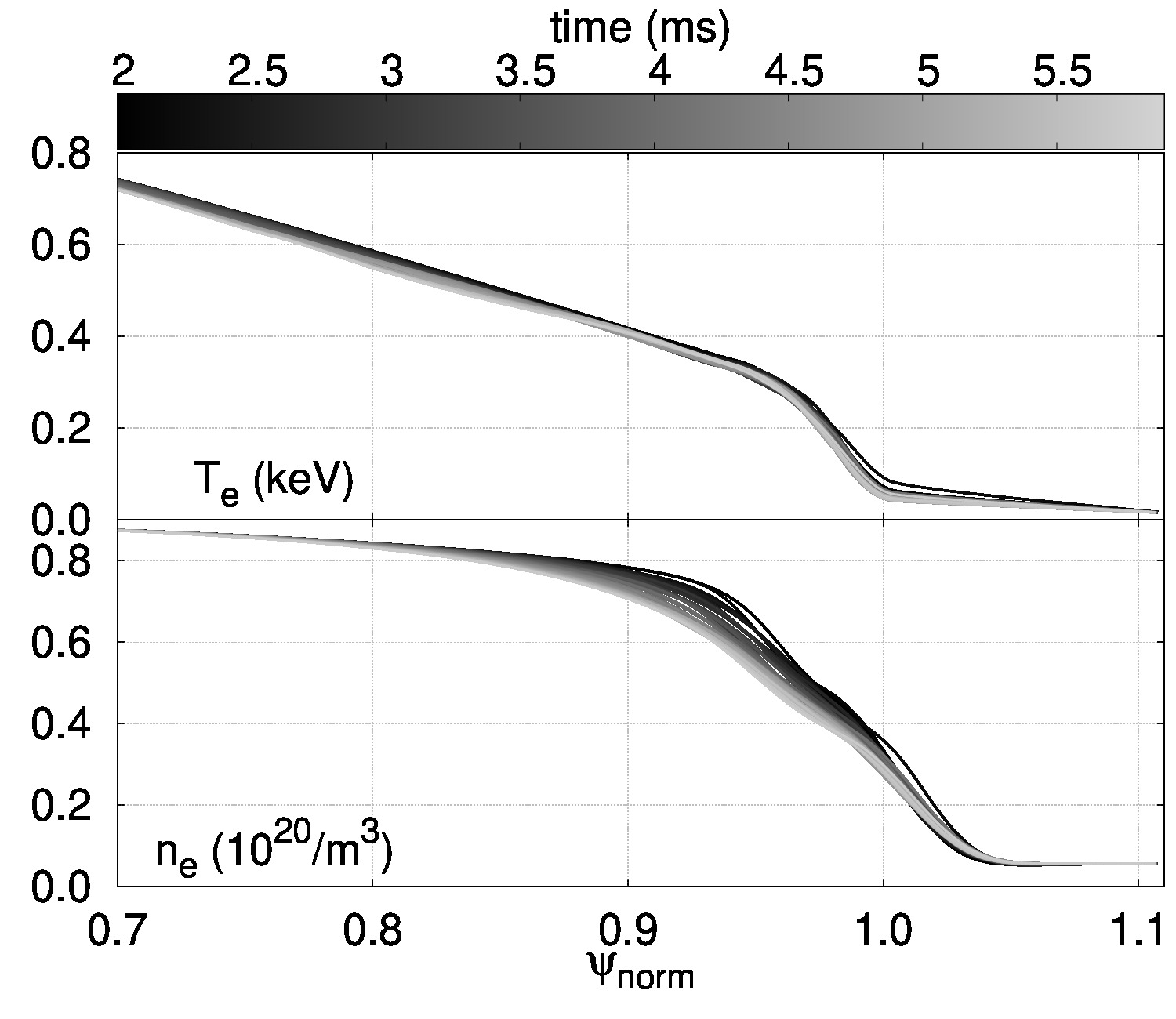}
\caption{Toroidally-averaged edge density and temperature profiles at the outboard midplane during different times of the non-linear phase. The steady depletion of the pedestal due to the edge non-axisymmetric perturbations can be clearly observed (particularly in the density profiles).}
\label{fig:base-nonlinear-profiles}
\end{figure}

As the non-linear phase gets underway, the interaction between background plasma and non-axisymmetric perturbations causes heat and particle transport in a continuous manner. This is in opposition to the transport caused by edge localised modes in JOREK simulations which, when studied at realistic resistivity and including diamagnetic effects and bootstrap current, feature transient transport~\cite{Cathey_2020}. The rate of heat and particle losses resulting from the non-axisymmetric perturbations correspond to ${\sim2.5~\mathrm{MJ/s}}$ and ${\sim6.9\times10^{21}~\mathrm{ions/s}}$, respectively. The evolution of edge density and temperature profiles (toroidally-averaged at the outboard midplane) during the non-linear phase is shown in fig.~\ref{fig:base-nonlinear-profiles}. This causes a steady depletion of the pedestal density, which may be partly responsible for the reduction in the dominant toroidal mode numbers during the non-linear phase. The temperature profile also decrease, but it is less affected because the heating power remains constant and the number of particles in the pedestal is constantly decreasing. The heat and particles that escape the magnetic confinement due to the mode activity are ultimately deposited into the simplified divertor targets. 

\subsection{Fluctuations across the pedestal}

\begin{figure}[!ht]
\centering
  \includegraphics[width=0.45\textwidth]{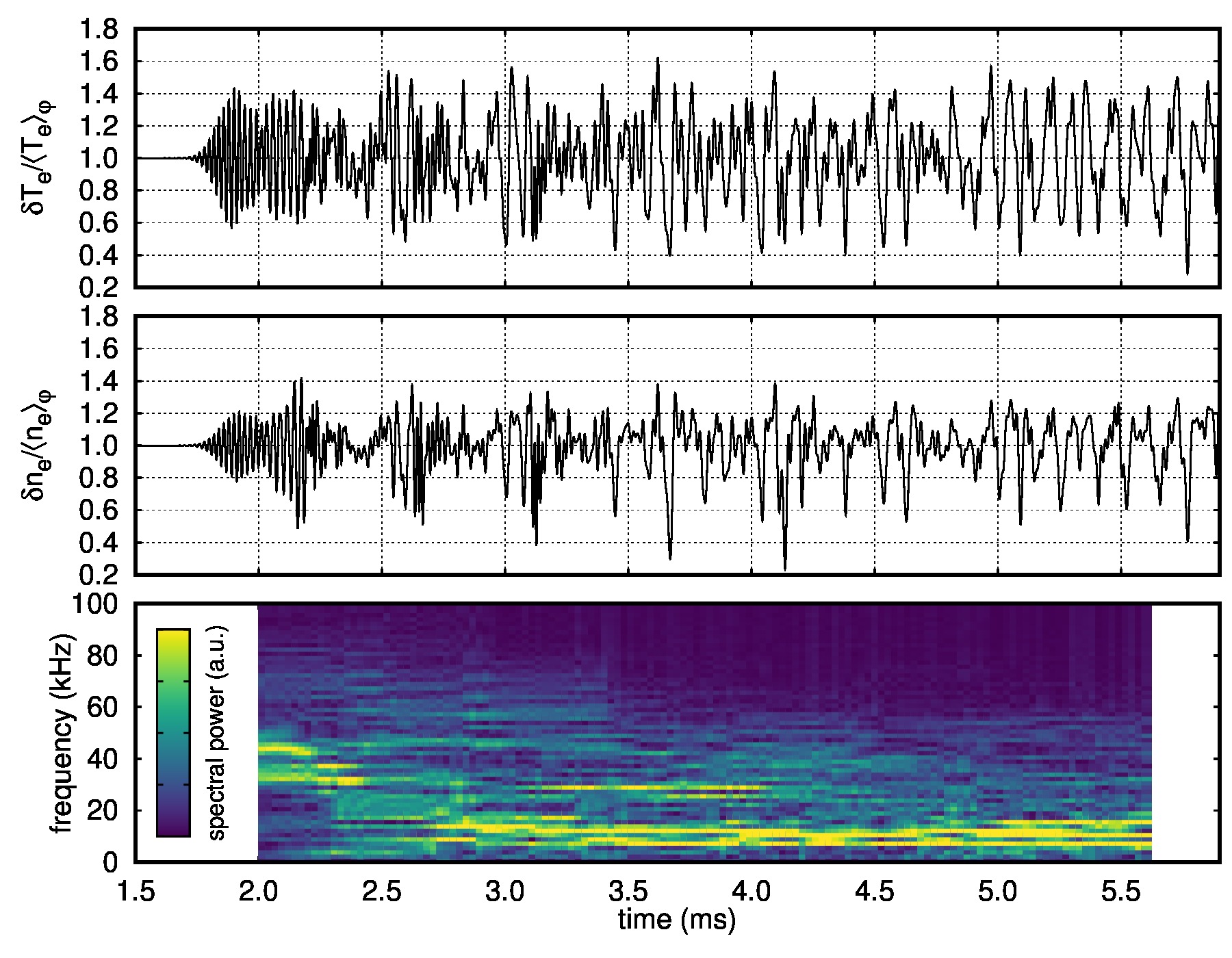}
\caption{Temperature (top) and density (middle) fluctuations relative to the toroidally-averaged value in the location where the mode amplitude is maximised (${\psi_\mathrm{norm}\approx0.98}$). Bottom: spectrogram of the temperature fluctuations. At the end of the early non-linear phase (${t\lesssim2.2~\mathrm{ms}}$) the pedestal fluctuations correspond to a frequency of roughly $35~\mathrm{kHz}$, but they chirp down to around ${13~\mathrm{kHz}}$ in the non-linear phase.}
\label{fig:base-nonlinear-spectrogram}
\end{figure}

Locally, the temperature and density fluctuations in the pedestal are very large, of around ${\delta T_e/\langle T_e \rangle \sim 50~\%}$ and ${\delta n_e/\langle n_e \rangle \sim 40~\%}$, which is consistent with observations of the QCM for both C-mod~\cite{LaBombard_2014} and AUG~\cite{Gil_2020}. This, together with a spectrogram of the temperature fluctuations signal is shown in fig.~\ref{fig:base-nonlinear-spectrogram}. In the spectrogram it is possible to determine that in the early non-linear phase (up to ${\sim2.2~\mathrm{ms}}$) the fluctuations cover the frequency range ${\{30\dots45\}~\mathrm{kHz}}$ and their frequency chirps down when the non-linear phase gets underway. This is due to the fact that the radial electric field well becomes less deep due to the interaction between non-axisymmetric perturbations and background plasma. During the non-linear phase, the frequency remains relatively constant in a frequency range of ${\{8\dots18\}~\mathrm{kHz}}$. On average, during this phase, the $E_r$ minimum is ${\sim-15~\mathrm{kV/m}}$, which is shallower than the experimental measurement (${-20\pm4~\mathrm{kV/m}}$). From the experimental measurements through helium beam (fig.~\ref{fig:aug36330-traces}), it is clear that the QCM frequency sits in the frequency range of ${f_\mathrm{QCM}\approx\{20\dots40\}~\mathrm{kHz}}$. This frequency range overlaps with what is obtained in the simulations during the linear phase (where the profiles have not yet experienced the depletion due to non-linear interactions with the perturbations). This is an indication that the resistive peeling-ballooning modes in the simulation could be related to the QCM. 

The poloidal structure of the non-axisymmetric perturbations (along a single flux-surface) features a predominantly ballooning structure. In the outer midplane during the non-linear phase, the range of poloidal wavenumbers is ${k_\theta\sim\{0.1\dots0.5\}~\mathrm{cm^{-1}}}$. Conversely, the reported poloidal wavenumbers for the QCM in C-mod are ${\sim1.5~\mathrm{cm^{-1}}}$. For AUG, based on measurement with helium beam the range of poloidal wavenumers for the QCM is ${\sim0.6}$ to ${0.9~\mathrm{cm^{-1}}}$, and for fluctuations during the quasi-continuous exhaust regime ${k_\theta\sim\{0.4\dots0.9\}~\mathrm{cm^{-1}}}$. As such, the poloidal wavenumbers hold a closer resemblance to the fluctuations measured during the QCE regime than the quasi-coherent mode during the EDA. 

\section{Discussion and conclusions}\label{sec:discussions-conclusions}

In this paper, first non-linear extended MHD simulations that consider initial conditions from an experimental discharge in ASDEX Upgrade that corresponds to the enhanced D-alpha H-mode (EDA) were presented. The equilibrium reconstruction was carried out with the CLISTE code~\cite{mc1999analytical} and linear ideal MHD stability analysis with MISHKA~\cite{Mikhailovskii_1997}, which found the experimental point inside the stable region near the high-n ballooning boundary. The experimental discharge under consideration features the EDA with a quasi-coherent mode (QCM) in a frequency range of 20 to ${40~\mathrm{kHz}}$ and is completely devoid of large edge localised modes. The cross-field transport that prevents the pedestal from building up is thus thought to be caused by the QCM. In Alcator C-mod, where the EDA was first observed, and routinely obtained, magnetic fluctuations were not observed in magnetic pick-up coils, but in ASDEX Upgrade, magnetic pick-up coils in the outboard midplane indeed observe fluctuations with toroidal mode numbers around ${n=\{3\dots7\}}$. The fundamental (i.e., $n=1$) frequency of these fluctuations in the discharge under consideration is ${32.9~\mathrm{kHz}}$. The relationship between these magnetic fluctuations and the QCM is not yet clear; the fundamental frequency in some discharges matches that of the QCM, while in others it displays a difference to ${f_\mathrm{QCM}}$ by up to $\sim15~\mathrm{kHz}$.

The simulations presented in this paper are carried out with the JOREK code~\cite{Hoelzl_2021,huysmans2007mhd}. The linearly most unstable modes are found to be resistive peeling-ballooning modes with toroidal mode numbers in the range ${n=\{6\dots9\}}$. The identification of such instabilities as resistive peeling-ballooning modes comes from their spatial structure (predominantly on the low-field side) and from their poloidal rotation velocity, which corresponds to that of resistive ballooning modes: ${v_\mathrm{ExB}+\bm v_\parallel\cdot \bm b_\theta}$~\cite{morales2016edge}. The location of the maximum mode amplitude corresponds to the maximum of the electron temperature gradient, and at this location the poloidal mode velocity is in the ion diamagnetic direction in the plasma frame; both such observations are in qualitative agreement with experimental measurements from gas puff imaging in Alcator C-mod~\cite{Theiler_2017}. As the linearly most unstable modes (notably $n=7$) grow in amplitude, they are able to non-linearly drive the linearly stable modes through three-wave interactions. As such, perturbations with lower and higher toroidal mode numbers (than ${n=4}$ and $9$, respectively) become excited. This early non-linear phase gives way to the fully non-linear phase when the non-axisymmetric perturbations grow large enough to interact with the background plasma. As the non-linear phase gets underway, the dominant toroidal mode number decreases from ${n=7}$ to ${n=3}$ but intermediate and high toroidal mode numbers remain relevant. During this phase, particles and heat are quasi-continuously expelled by the cross-field transport generated by the non-axisymmetric perturbations. As such, the pedestal density starts to become depleted. 

The density and temperature fluctuations during the non-linear phase have large amplitudes, which is qualitatively consistent with measurements from EDA in ASDEX Upgrade~\cite{Gil_2020}. Before the early non-linear phase comes to an end, these fluctuations lie on a frequency range of ${\{30\dots40\}~\mathrm{kHz}}$, which is consistent with the observations for the present discharge (before the density and temperature profiles are affected by the non-linear interaction between the non-axisymmetric perturbations and the background plasma). During the fully non-linear phase, when the density pedestal is depleting and the radial electric field well is becoming shallower, the fluctuation frequencies decrease to ${\{8\dots18\}~\mathrm{kHz}}$ (which still sits at the lower end of the observed values of ${f_\mathrm{QCM}}$ in AUG~\cite{Gil_2020}).

In summary, from the analysis performed of the JOREK simulations of an EDA H-mode, several quantitative and qualitative similarities are found with respect to experimental features and measurements of the quasi-coherent mode and magnetic fluctuations captured by the pick-up coils. Further comparisons between non-linear extended MHD simulations and experimental measurements from the EDA H-mode are left for future work. In particular, a focus on more quantitative comparisons of the EDA H-mode itself and of its boundaries, including its relation to the quasi-continuous exhaust (QCE) regime is foreseen. To this end, taking into account separate electron and ion temperatures, resistive walls instead of the ideal wall boundary condition employed here, advanced SOL models, toroidal rotation sources, and potentially higher order finite Larmor radius corrections are under consideration. Additional dependencies on the viscosity values, heat and particle sources, and impurity radiation can additionally be taken into account.

\section*{Acknowledgements}

This work has been carried out within the framework of the EUROfusion Consortium, funded by the European Union via the Euratom Research and Training Programme (Grant Agreement No 101052200 — EUROfusion). Views and opinions expressed are however those of the author(s) only and do not necessarily reflect those of the European Union or the European Commission. Neither the European Union nor the European
Commission can be held responsible for them. In particular, A. Cathey received funding from a EUROfusion Researcher Grant and work package Tokamak Exploitation, and are acknowledged. The non-linear JOREK simulations were performed using the Marconi-Fusion supercomputer within the FUA36\_MHD project.

\section*{References}

\bibliography{references}

\end{document}